\documentclass[aps,multicol,epsfig,twocolumn,showpacs]{revtex4}
\newcommand\be{\begin{eqnarray}}
\newcommand\ee{\end{eqnarray}}
\newcommand\ba{\begin{array}}
\newcommand\ea{\end{array}}
\def\r{\rangle}
\def\l{\langle}

\def\T{{\rm Tr}}
\def\cH{{\cal H}}
\def\cS{{\cal S}}

\def\cI{{\cal I}}

\def\cE{{\cal E}}

\def\openone{{\it I}}

\usepackage{graphicx}
\begin{document}
\title{Entanglement-induced state ordering under local operations}
\author{M\'ario Ziman$^{1,2,3}$ and Vladim\'\i r Bu\v zek$^{1,3,4}$}
\address{
$^1$Research Center for Quantum Information, Slovak Academy of Sciences,
D\'ubravsk\'a cesta 9, 84511 Bratislava, Slovakia
\\
$^2$Faculty of Informatics, Masaryk University, Botanick\'a 68a, 60200
Brno, Czech Republic\\
$^3$Quniverse, L\'\i\v s\v cie \'udolie 116, 841 04 Bratislava, Slovakia\\
$^4$Abteilung f\"{u}r Quantenphysik, Universit\"at Ulm, 89069 Ulm,
Germany }
\begin{abstract}
We analyze how entanglement between two components of a bipartite system behaves under the action of local channels
of the form $\cE\otimes\cI$. We show that a set of maximally entangled states is by the action
of $\cE\otimes\cI$ transformed into the set of states that exhibit the same degree of entanglement.
Moreover, this degree represents an upper bound on entanglement that is available at the output of the channel
irrespective what is the input state of the composite system. We  show that within this bound the
the entanglement-induced state ordering is ``relative'' and can be changed by the action of local channels. That is, two states
$\varrho_1^{(in)}$ and $\varrho_2^{(in)}$
such that the entanglement $E[\varrho_1^{(in)}]$ of the first state is larger than the entanglement $E[\varrho_2^{(in)}]$
of the second state are transformed into states  $\varrho_1^{(out)}$ and $\varrho_2^{(out)}$
such that $E[\varrho_2^{(out)}] > E[\varrho_1^{(out)}]$.
\end{abstract}

\maketitle
The success of quantum information theory
\cite{nielsen,presskill} is intimately related to the phenomenon
of quantum entanglement. The better we will understand properties
of this purely quantum phenomenon the deeper will be our insight
into the quantum realm. Even though the importance of quantum
entanglement has been clearly acknowledged by founding fathers of
quantum mechanics \cite{schrodinger}, the true potential of this
phenomenon has been appreciated just recently with the development
of quantum information science. Over last ten years many results
of fundamental importance illuminating properties of quantum
entanglement have been reported. In spite of all the progress,
there are still many questions that are to be answered. In particular,
criteria of non-separability of arbitrarily-dimensional bi-partite
systems, the study of intrinsic multi-partite
entanglement in composite quantum systems
\cite{wootters_ckw,osborne}, or the role of quantum
entanglement in macroscopic systems is presently under
investigation \cite{vedral}. One of the problems that has
attracted interest of researchers for quite some time is the issue
of ``proper'' measures of entanglement \cite{plenio}.

In general, we can identify two conceptually different approaches in various
attempt to define measures of entanglement. These can be named as
{\it i)} the operational approach, and {\it ii)} the formal (abstract)
approach. The operational approach is based on an assumption that there exists
a process, or an
information protocol, in which the quantum entanglement plays the role of a new
resource that provides some improvement in the performance of the protocol  compared to its
``classical analogue''. The second approach \cite{plenio} is based on
postulation of the desired properties that an entanglement measure
has to satisfy and defines a functional with these properties (see below).

In this paper we will adopt this second approach. We will start our discussion
by addressing a specific question concerning the
entanglement measures. First, we will answer the question: {\it Do local operations preserve the
entanglement-induced ordering?} Then we will analyze how entangled states  that exhibit the same degree of entanglement
are transformed under the action
of local channels.

Let us start with a trivial observation: States of bi-partite quantum systems can be either entangled or separable
(this is almost a tautological statement since the presence of entanglement is defined as an absence of the separability,
and {\it vice versa}).
On the other hand, the definition of entanglement is related to the non-existence of local
hidden variable model for an observed statistics for a given state $\varrho^{AB}$ of a bi-partite system.
R.F.~Werner in his seminal work \cite{werner} has shown that such assumption restricts the entangled
states to those that cannot be written as convex combinations of product
states, i.e. $\varrho\ne \sum_k p_k \varrho_k^A\otimes \varrho^B_k$. The
entanglement measure $E$ is a positive functional defined on the state
space of a bi-partite quantum system. Following Plenio
and Vedral \cite{plenio},
let us summarize basic properties that any
entanglement measure has to satisfy:
\begin{enumerate}
\item{\it Sharpness:} $E(\varrho)=0$ if and only if $\varrho^{AB}$ is separable.
\item{\it Local unitary invariance:} $E(\varrho^{AB})=E(U_A\otimes U_B\varrho^{AB} U_A^\dagger\otimes U_B^\dagger)$.
\item{\it Convexity:} $E(\sum_k p_k \omega^{AB}_k)\le \sum_k p_k E(\omega^{AB}_k)$
\item{\it Normalization:} $E(\varrho^{AB})=\max_\varrho E(\varrho^{AB})$
if and only if $\varrho^{AB}=\left(\varrho^{AB}\right)^2$,
$\T_A\varrho^{AB}=\T_B\varrho^{AB}=\frac{1}{2}\openone$. States with such properties are
called as {\it the maximally entangled states}.
\item{\it Non-increasing under local operations and classical communication (LOCC):} A general LOCC transforms the original state
$\varrho^{AB}$ into
a mixture of states $\omega_k^{AB}$ with a probability $p_k$. We require
that on average the entanglement cannot be increased, i.e.
$\sum_k p_k E(\omega_k^{AB})\le E(\varrho^{AB})$. Let us note that
$\varrho_k^{AB} = \cE_k^A\otimes \cE_k^B[\varrho^{AB}]$.
\item{\it Pure state additivity:}
$E(\Phi_1\otimes\Phi_2)=E(\Phi_1)+E(\Phi_2)$
for all pure states $\Phi_1,\Phi_2$.
\end{enumerate}

For all protocols in which quantum entanglement serves as a resource it is true
that higher the degree of entanglement more efficient the application of the protocol is.
However, even for a two-qubit system there exist several
 measures of entanglement satisfying the above properties (e.g. the concurrence \cite{wootters},
the relative entropy of entanglement \cite{henderson}, etc.) and nevertheless
each of these measures might induce a different ordering
on the set of states \cite{eisert,grudka}. Thus, it is difficult to say ``objectively'' which
states are more entangled. The maximally entangled states play a
special role here, because the normalization property single them out
independently of the particular measure.

Recently several authors
\cite{zyckowski,verstraete,cai,roszak,cardoso,liang,ziman_cepe}
have investigated time evolution of the entanglement when a bi-partite system has been subjected
to either local or global transformations (time evolution).
In these studies, usually it has been investigated how
a {\em maximally} entangled state is affected by the action of the
corresponding evolution operation. Certainly, it is very instructive to
understand what happens
 if the bi-partite system is initially prepared in a non-maximally entangled state
$\varrho$ (in order to simplify the notation, in what follows we will omit superscripts
indicating that the density operator describes two systems $A$ and $B$).
To illuminate this problem from a perspective of an action of local channels let us formulate the following theorem.

\noindent\newline
{\bf Theorem.}
{\it For each state $\omega\in\cS(\cH\otimes\cH)$ and a quantum
local channel $\cE_L=\cE\otimes\cI$ the following inequality holds
\be
E(\cE_L[\omega])\le E(\cE_L[\Psi_+])\; ,
\ee
where $E$ is some (normalized) entanglement measure, i.e.
$\max_\omega E(\omega)=E(\Psi_+)$, where
$\Psi_+$ is a maximally entangled state.}

\noindent\newline
{\it Proof.} Let us assume the convexity of
the entanglement measure, i.e. $E(\omega)\le \sum_j q_j
E(\omega_j)$. Under such assumption it is sufficient to consider only
pure states, i.e. to show that $E(\cE_L[\Phi])\le E(\cE_L[\Psi_+])$.
Any pure bipartite state $\Phi=|\phi\r\l\phi|$ can be written as
$|\phi\r=I\otimes A|\psi_+\r$,  where $A$ is a suitable linear
operator. Hence we have $\cE_L[\Phi]=(\cE\otimes\cI)[(I\otimes A)\Psi_+
  (I\otimes A^\dagger)]=(I\otimes A)(\cE\otimes\cI)[\Psi_+](I\otimes
A^\dagger)=\cE_A\cE_L[\Psi_+]$. Next we will use the fact that local
actions cannot increase the value of entanglement, i.e. $E(\cE_A[\rho])\le
E(\rho)$. Consequently, by putting $\rho=\cE_L[\Psi_+]$ we obtain the
desired inequality $E(\cE_L[\Phi])\le E(\cE_L[\Psi_+])$.

As a consequence we have that a local action applied on
a set of maximally entangled states results in a set of states with the same
amount of entanglement (irrespective of the measure we use).
One can prove this property directly from
the fact that all
maximally entangled states are related by local unitary transformation applied
on one subsystem, i.e. the states
$|\Psi_U\r=I\otimes U |\Psi_+\r$ form a set of
all maximally entangled states. Applying the local operation
$\cE\otimes I[\Psi_U]=\cE\otimes U [\Psi_+]
=(\cI\otimes U)(\cE\otimes\cI[\Psi_+])=(\cI\otimes U)[\Omega_\cE]$.
This means that states $\Omega_\cE=\cE\otimes \cI[\Psi_+]$ and
$\Omega_\cE^U = \cE\otimes\cI[\Psi_U]$ are locally unitary equivalent
($\Omega_\cE^U=\cI\otimes U[\Omega_\cE]$).
Consequently, they contain the same amount of entanglement, which proves
our statement. Such result can be extended to all
states $\omega_1,\omega_2$ equivalent in the sense
$\omega_1=\cI\otimes U[\omega_2]$, for which
the equality of entanglement is not affected by the local action
$\cE\otimes\cI$.

The above theorem has served as the motivation
to the main question considered in this paper.
Specifically, whether all local channels of the form $\cE_L=\cE\otimes\cI$
preserve the ordering induced by a given entanglement measure.
Or in other words, for which entanglement measures $E$ the following
implication holds
\be
E(\omega_1)\le E(\omega_2)\Rightarrow E(\omega_1^\prime)
\le E(\omega_2^\prime)\, ,
\ee
where $\omega_j^\prime=\cE\otimes\cI[\omega_j]$ ($j=1,2$).
One might think that for a ``good'' measure of entanglement
such inequality should be valid. However, this property is
not listed among the usually required properties of entanglement
measures. Surprisingly enough,
there is a simple argument that such ``natural''
property cannot hold in general. Firstly, we will
present an explicit counterexample for the entanglement measure
of two-qubit states called the concurrence \cite{wootters}. Secondly, we will give
a general argument why the answer to our question is negative,
i.e. local actions may affect nontrivially
the ordering induced by entanglement measures.

\noindent\newline
{\bf Example 1.}
Let us consider two families of two-qubit states: {\it i)} Werner states
$\varrho_1=q\Psi_+ + (1-q)\frac{1}{4}I$ ($0\le q\le 1$),
and {\it ii)} pure states $\varrho_{2}=\Phi=|\phi\r\l\phi|$,
$|\phi\r=\alpha|00\r+\beta|11\r$ with $\alpha,\beta$ real and
$\alpha^2+\beta^2=1$. Both of these families of states
cover the whole interval of possible values of entanglement.
Applying the local depolarizing channel
$\cE[\varrho]=p\varrho+(1-p)\frac{1}{2}I$ we obtain
\be
\varrho_1^{\rm in} &\stackrel{\cE\otimes\cI}{\longrightarrow}&
\varrho_1^{\rm out} = pq\Psi_+ + (1-pq)\frac{1}{4}I\; ; \\
\varrho_{2}^{\rm in} &\stackrel{\cE\otimes\cI}{\longrightarrow}&
\varrho_{2}^{\rm out}=p\Phi+(1-p)\frac{1}{2}\openone\otimes \varrho_B\; ,
\ee
where $\varrho_B=\T_A\Phi$. The states
($\varrho_1^{\rm in},\varrho_1^{\rm out},
\varrho_{2}^{\rm in},\varrho_{2}^{\rm out}$)
belong to the set of states represented by $\varrho_{2}^{\rm out}$,
i.e. a mixture of pure state $\Phi$ with the mixed state
$\frac{1}{2}\openone\otimes\varrho_B$. Therefore, it is
sufficient to calculate the amount of entanglement for the state
$\varrho_{2}^{\rm out}$
\be
\varrho_{2}^{\rm out}=\left(
\ba{cccc}
\alpha^2\frac{1+p}{2} & 0 & 0 & p\alpha\beta \\
0 & \beta^2\frac{1-p}{2} & 0 & 0 \\
0 & 0 & \alpha^2\frac{1-p}{2} & 0 \\
p\alpha\beta & 0 & 0 & \beta^2\frac{1+p}{2} \\
\ea
\right)\, .
\ee

\begin{figure}
\includegraphics[width=8cm]{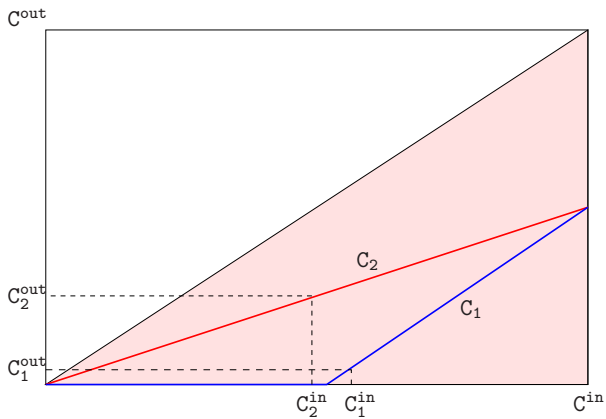}
\caption{(Color online)  The input/output concurrence diagram for
both families of states $\varrho_1,\varrho_2$ and for the
depolarizing channel with $p=1/2$. The states from the
counterexample 1 discussed in the paper are displayed and the
change in the ordering is visible. The region under the line
$C^{\rm out}=C^{\rm in}$ represents the allowed region that is
achievable by local channels. Concurrence is measured in
dimensionless units. \label{fig} }
\end{figure}

The entanglement measure called the concurrence \cite{wootters} is defined
via the square roots of  eigenvalues of the matrix
$R=\varrho_{2}^{\rm out}(\sigma_y\otimes\sigma_y)
[\varrho_{2}^{\rm out}]^*(\sigma_y\otimes\sigma_y)$.
Let us denote by
${\rm Eig}[R]=\{\lambda_1,\lambda_2,\lambda_3,\lambda_4\}$
the eigenvalues of the matrix $R$ ordered in a decreasing order,
i.e. $\lambda_1\ge \lambda_2\ge\lambda_3\ge\lambda_4$.
The concurrence is given by the expression
$C=\max\{0,\sqrt{\lambda_1}-\sqrt{\lambda_2}
-\sqrt{\lambda_3}-\sqrt{\lambda_4}\}$.
For the states
$\varrho_{2}^{\rm out}$ one can find a compact expression for the concurrence
$C_2^{\rm out}=\alpha\beta(3p-1)$. Using this formula we find
all quantities that are need for our considerations. In particular,
\be
C_1^{\rm in}=\frac{1}{2}(3q-1) & \stackrel{\cE\otimes\cI}{\longrightarrow} &
C_1^{\rm out}= \frac{1}{2}(3pq-1)\; ; \\
C_2^{\rm in}=2\alpha\beta \ \ \ \ & \stackrel{\cE\otimes\cI}{\longrightarrow} &
C_2^{\rm out}= \alpha\beta(3p-1)\; .
\ee
Moreover, we are able to determine the functional dependence
$C^{\rm out}=f(C^{\rm in})$. In particular, we obtain parametric
linear functions
\be
\nonumber
\varrho_1 &:& C^{\rm out}_1=\max\{0,p C^{\rm in}_1 +\frac{1}{2}(p-1)\}\, ; \\
\varrho_2 &:& C^{\rm out}_2=\max\{0,\frac{3p-1}{2} C^{\rm in}_2\}\, ,
\ee
where the parameter $p$ represents the action (strength) of the depolarizing channel.
Let us consider the depolarizing channel with $p=0.5$ and two input states such that
$C_1^{\rm in}=\frac{1}{2}+\epsilon$, $C_2^{\rm in}=\frac{1}{2}-\epsilon$.
The degree of entanglement
of the resulting output states reads $C_1^{\rm out}=\frac{1}{2}\epsilon$ and
$C_2^{\rm out}=\frac{1}{8}-\frac{1}{4}\epsilon$. It is easy to see that
whenever $0<\epsilon < \frac{1}{6}$ the original inequality
$C_1^{\rm in}>C_2^{\rm in}$ is transformed  into the opposite inequality, i.e. $C_1^{\rm out}<C_2^{\rm out}$
{[see Fig.~\ref{fig}]}.

\noindent\newline
{\bf Example 2.}
Let us assume a four-qubit system divided into two pairs: the left (L)
and the right (R). Consider two pure states
$|\Omega_j\r=|\psi_j\r_{L_1 R_1}\otimes |\phi_j\r_{L_2 R_2}$ for $j=1,2$, i.e.
the parties L and R share two pairs of two-qubit pure states.
We will act locally on the right couple of qubits only, i.e.
we apply the channel $\cE=\cI_{L_1 L_2}\otimes \cE_{R_1 R_2}$. In particular,
we assume that $\cE_{R_1 R_2}=\cE\otimes\cI$ and $\cE$ is the contraction
into a fixed pure state, i.e. $\varrho\to\Xi$.
The additivity of entanglement for pure states implies that initially
$E(\Omega_1)=E(\Psi_1)+E(\Phi_1)=x_1+y_1$ and
$E(\Omega_2)=E(\Psi_2)+E(\Phi_2)=x_2+y_2$. After the action of the
local channel we obtain the states
$\Omega_1^\prime=[\T_{R_1}\Psi_1]\otimes\Xi_{R_1}\otimes\Phi_1$ and
$\Omega_2^\prime=[\T_{R_1}\Psi_2]\otimes\Xi_{R_1}\otimes\Phi_2$.
The convexity implies
$E(\Omega_1^\prime)\le \lambda
E(|\omega_1\r\l\omega_1|\otimes \Xi\otimes\Phi_1)
+(1-\lambda)E(|\omega_2\r\l\omega_2|\otimes \Xi\otimes\Psi_1)$, where
$|\omega_j\r$ are eigenvectors of $\T_{R_1}\Psi_1$ and $\lambda,1-\lambda$
are the corresponding eigenvalues. On the right hand side we have pure
states, for which the entanglement can be found by using the
additivity of $E$ for pure states. We obtain
$E(\Omega_1^\prime)\le \lambda E(\Phi_1)+(1-\lambda) E(\Phi_1)=E(\Phi_1)=y_1$
and by a similar line of arguments we find $E(\Omega_2^\prime)\le y_2$. Let us consider
that originally $x_1+y_1 > x_2+y_2$ ($E(\Omega_1)>E(\Omega_2)$)
and $y_2 > y_1=0$. After the action of the local channel
we have $E(\Omega_1^\prime)=0$ and $E(\Omega_2^\prime)\le y_2$, i.e.
we obtain an un-sharp inequality $E(\Omega_1^\prime)\le E(\Omega_2^\prime)$.
In what follows we will argue that this inequality cannot be saturated, i.e.
$0<E(\Omega_2^\prime)$. In other words, we would like to show that
there is still some entanglement present between the left and the right systems.
Since one pair ($\Phi_1$, or $\Phi_2$)
has not been affected at all by the action of the local channel it can be still used to violate
Bell inequalities (the state $\Phi_2$ is pure and entangled). This violation of Bell inequalities  is
an evidence of presence of entanglement between the two parties. In conclusion,
$E(\Omega_1^\prime)<E(\Omega_2^\prime)$, but $E(\Omega_1)>E(\Omega_2)$.


Let us summarize the line of arguments in the last example:
We have two pairs of qubits such that $x_1+y_1>x_2+y_2$ and
we put $y_1=x_2=0$, i.e. $x_1>y_2$. In other words we have two different
pure states of two pairs: 1) the entangled$\otimes$factorized, and
2) the factorized$\otimes$entangled states. The contractive channel is applied on the first
pair only. The first state is transformed and become unentangled,
while the second state can be used to violate  Bell inequalities, because
the entangled part has not been affected at all by the action of the local channel.
As a result, we conclude that the entanglement-induced
ordering of states is not absolute
under the action of local channels. Consequently, in some circumstances
less entangled states can be more ``robust'' against local operations.

Using the above theorem we can justify the choice of maximally
entangled states in our analysis of time-dependent entanglement ordering induced
by local evolutions. That is, maximally entangled states when subjected to a local evolution
(i.e., one particle is evolving freely while the second is subjected to an action of a local channel)
bound the maximally available entanglement. However,
one has to be careful to draw
general conclusions about the entanglement behavior.
Here we have analyzed the simplest situation in which only one of
the subsystems undergone a nontrivial local evolution.
Let us note that the Theorem does not hold if one considers bi-local
channels, i.e. in that case even the maximality of entanglement
is a ``relative'' notion.

The second example illustrates that
the entanglement-induced ordering is not absolute in general: The original
entanglement-based ordering (irrespective of the choice of the entanglement measure)
within a set of states can be changed (even) under the action of a local channel.
However, for two-qubit systems it is still an open question,
whether there exist an entanglement measure, for which the entanglement-induced
state ordering is not affected by the action of local channels.
Another open problem is the characterization
of those local channels that preserve the state ordering induced
by a given measure, for instance the concurrence.
The characterization and classification of
channels using the
$E^{\rm out}$ {\it vs.} $E^{\rm in}$ diagram can be very useful
in such analysis.  This approach can be applied to
any measure and any channel.
For instance, the action of local
unitary channels $U\otimes\cI$ are represented by the line $E^{\rm out}=E^{\rm in}$
in this diagram. Analogously, for entanglement-breaking
channels \cite{eb_channels,eb_horodecki} we have $E^{\rm out}=0$,
i.e. the action of these channels is represented again by a line in the $E^{\rm out}$ {\it vs} $E^{\rm in}$ diagram.
Both of these families of channels preserve the entanglement-induced
ordering for an arbitrary measure of entanglement. Presently it is still an open problem
whether there exist some other local channels that have the same property
(at least for a particular measure of entanglement).

From the construction that has been used in Example 1 it follows that
whenever we obtain a region that is bounded by two lines corresponding to two input states
that exhibit the same degree of initial entanglement
in the $E^{\rm out}$ {\it vs} $E^{\rm in}$ diagram
we can find two states for which the entanglement-induced ordering
is not preserved. Channels represented by a line
in $E^{\rm out}$ {\it vs.} $E^{\rm in}$ diagram
not only preserve the state ordering, but they also preserve
the ``isoentangled'' classes of states. In other words,
if the ``in'' states are equally entangled, then
the ``out'' states are equally entangled as well.
This is a very confining constraint imposed on the action of local quantum channels and
therefore we do not expect that there exists a nontrivial channel (except unitary, or
entanglement-breaking channels mentioned above) that
preserve the entanglement-induced ordering.

In conclusion, we have addressed the question how the
entanglement between two components of a bipartite system behaves under the action of local channels
of the form $\cE\otimes\cI$. We have shown that a set of maximally entangled states is by the action
of the local channel $\cE\otimes\cI$ transformed into the set of states that exhibit the same degree of entanglement.
Moreover, this degree represents an upper bound of entanglement that can be available at the output of the channel
irrespective what is the input state. We have shown that within this bound the
 the entanglement-induced state ordering is ``relative'' and can be changed by the action of local channels.
Moreover, our study suggests that this is a rather common property of most of the local channels. That is, most of
the local channels affect the entanglement-induced state ordering.
Our analysis opens several
interesting questions related to ``dynamics'' of
entanglement and provides a novel tool
of the characterization of arbitrary quantum operations
(both local and global) using the $E^{\rm out}$ {\it vs.} $E^{\rm in}$ diagram.
Let us note that already for the so called bi-local channels of
 the form $\cE_1\otimes\cE_2$ this picture
looks differently, because in this case even the set of maximally of
entangled states transforms differently, i.e. the output states
might not posses the same degree of entanglement.
In addition, it is not clear whether the ``maximality''
of entanglement is absolute, or not, i.e. whether
$\omega_{\max}={\rm arg}\sup_\omega E(\cE_1\otimes\cE_2[\omega])$ is
one of the maximally entangled states, or not.
This problem illustrates the fact
that dynamical features of quantum entanglement
are still not understand in all details and many interesting features
remain to be explored and understood.

\noindent{\bf Acknowledgments}\newline
This work has been supported partially
by European projects QGATES, CONQUEST,
by Slovak project APVT
and by GA\v CR GA201/01/0413. We also acknowledge the
support of Slovak Academy of Sciences via
the project CE-PI (I/2/2005) and VB would like to thank to the Alexander von Humboldt
Foundation.




\begin{thebibliography}{00}
\bibitem{eb_channels}
The channel is called as the entanglement-breaking if
the state $\varrho^\prime=\cE\otimes\cI[\varrho]$
is separable for all $\varrho$.
\end{thebibliography}
\end{document}